\documentclass[prb,twocolumn,showpacs]{revtex4}
\usepackage{graphicx} 
\usepackage{amsmath,amsthm,amssymb,mathrsfs}
\usepackage{bm}

\begin{document}

\title{Magnetoresistance generated from charge-spin conversion by anomalous Hall effect in metallic ferromagnetic/nonmagnetic bilayers}

\author{Tomohiro Taniguchi}
 \affiliation{
 National Institute of Advanced Industrial Science and Technology (AIST), Spintronics Research Center, Tsukuba, Ibaraki 305-8568, Japan 
 }

 \begin{abstract}
{ 
A theoretical formulation of magnetoresistance effect in a metallic ferromagnetic/nonmagnetic bilayer 
originated from the charge-spin conversion by the anomalous Hall effect is presented. 
Analytical expressions of the longitudinal and transverse resistivities in both nonmagnet and ferromagnet are obtained by solving the spin diffusion equation. 
The magnetoresistance generated from charge-spin conversion purely caused by the anomalous Hall effect in the ferromagnet is 
found to be proportional to the square of the spin polarizations in the ferromagnet and has fixed sign. 
We also find additional magnetoresistances in both nonmagnet and ferromagnet 
arising from the mixing of the spin Hall and anomalous Hall effects. 
The sign of this mixing resistance depends on those of the spin Hall angle in the nonmagnet 
and the spin polarizations of the ferromagnet. 
}
 \end{abstract}

 \maketitle




\section{Introduction}
\label{sec:Introduction}

It was recently found that the resistance of a bilayer consisting of 
an insulating ferromagnet and a metallic nonmagnet depends on the magnetization direction 
even though no current flows in the ferromagnet [\onlinecite{weiler12,huang12,nakayama13,althammer13,hahn13}]. 
This new type of magnetoresistance effect, called spin Hall magnetoresistance, 
was explained theoretically by using the diffusive spin transport theory in the nonmagnet [\onlinecite{chen13}], 
despite the existence of other contributions being implied [\onlinecite{lu13,miao14,li14}]. 
The key idea of the spin Hall magnetoresistance was the charge-spin conversion [\onlinecite{zhang00,dyakonov07,takahashi08,ando08,liu11,liu12,haney13,saslow15}] 
caused by the spin Hall effect [\onlinecite{dyakonov71,hirsch99,hoffmann13}] originating from the spin-orbit interaction in nonmagnetic heavy metals. 
The spin Hall magnetoresistance was observed also in a metallic ferromagnetic/nonmagnetic bilayer [\onlinecite{avci15,liu15,cho15,kim16}]. 


A metallic ferromagnet shows the anomalous Hall effect [\onlinecite{pugh53}] 
generating another electric voltage in the direction perpendicular to both the magnetization and an external electric field. 
Note that the transverse electric current generated by this Hall voltage is spin-polarized 
because of the spin-dependent transport properties in ferromagnets [\onlinecite{miao13,taniguchi15}]. 
In other words, the anomalous Hall effect generates the electric and spin currents simultaneously. 
Then, a new magnetoresistance effect that originates from the charge-spin conversion by the anomalous Hall effect, 
in addition to this conventional Hall voltage, is expected, 
as in the case of the spin Hall magnetoresistance. 
This new type of magnetoresistance effect can be distinguished from the conventional anomalous Hall effect 
because the former is characterized by the physical quantities related to the spin-dependent transport in a ferromagnet. 


In this paper, a theoretical formulation is given for the magnetoresistance effect generated from 
the charge-spin conversion caused by the anomalous Hall effect in a ferromagnetic/nonmagnetic bilayer. 
Solving the spin diffusion equation and using the spin-dependent Landauer formula, 
analytical expressions of the transverse resistivities in both nonmagnet and ferromagnet are obtained, 
both of which have angular dependence that is the same as with the planar Hall effect. 
One of these magnetoresistances originates from the charge-spin conversion caused purely by the anomalous Hall effect in the ferromagnet. 
This contribution has fixed sign because it is proportional to the square of the spin polarizations of the ferromagnet. 
Another magnetoresistance effect arises from the mixing of the spin Hall and anomalous Hall effects. 
The sign of this term depends on those of the spin Hall angle in the nonmagnet and the spin polarizations of the ferromagnet. 
The longitudinal resistivity is also investigated and found to also has the mixing term. 


The paper is organized as follows. 
The electric and spin currents, as well as the spin accumulation, 
in a ferromagnetic/nonmagnetic bilayer are calculated 
in Sec. \ref{sec:Charge and spin transport in a ferromagnetic/nonmagnetic bilayer} 
by solving the spin diffusion equation. 
The magnetoresistance effect due to the charge-spin conversion 
by the spin Hall and anomalous Hall effects is studied in Sec. \ref{sec:Magnetoresistance effect due to charge-spin conversion}. 
The conclusion is summarized in Sec. \ref{sec:Conclusion}. 



\begin{figure}
\centerline{\includegraphics[width=1.0\columnwidth]{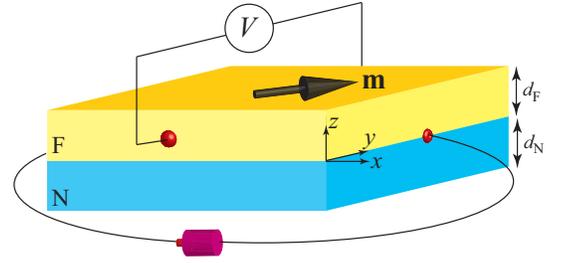}}
\caption{
         Schematic view of the system under consideration. 
         An external voltage is applied along the $x$ direction, generating the external electric field $E_{x}$. 
         Magnetoresistance in the transverse direction is measured from the electric voltage flowing along the $y$ direction in the nonmagnet or ferromagnet, 
         whereas the longitudinal one is detected from the electric voltage along the $x$ direction. 
         \vspace{-3ex}}
\label{fig:fig1}
\end{figure}



\section{Charge and spin transport in a ferromagnetic/nonmagnetic bilayer}
\label{sec:Charge and spin transport in a ferromagnetic/nonmagnetic bilayer}

In this section, we derive the analytical formulas of 
the electric and spin currents, as well as the spin accumulation, in a ferromagnetic/nonmagnetic bilayer. 


\subsection{Electric current}

The system we consider is schematically shown in Fig. \ref{fig:fig1}, 
where the ferromagnet (F) and nonmagnet (N) are attached at the $xy$-plane, $z=0$. 
In the following, we use the subscripts F and N to distinguish between the quantities related to the ferromagnet and nonmagnet. 
The thicknesses of the ferromagnet and nonmagnet along the $z$-direction are denoted as $d_{\rm F}$ and $d_{\rm N}$, respectively. 
The unit vector pointing in the magnetization direction of the ferromagnet is $\mathbf{m}$. 
An external voltage is applied along the $x$-direction. 
The electric current density in the nonmagnet is driven by the external electric field $E_{x}$ 
and the inverse spin Hall effect, 
and is given by [\onlinecite{takahashi08}]
\begin{equation}
  J_{{\rm c}i,{\rm N}}
  =
  \frac{\sigma_{\rm N}}{e}
  \partial_{i}
  \bar{\mu}_{\rm N}
  -
  \frac{\vartheta \sigma_{\rm N}}{e}
  \epsilon_{ij \alpha}
  \partial_{j}
  \delta
  \mu_{{\rm N},\alpha},
  \label{eq:current_N}
\end{equation}
where $i=x,y,z$ represents the spatial direction of the current flow, 
whereas $\alpha$ is used to represent the direction of the spin polarization. 
The spin-polarization direction of the spin accumulation in the nonmagnet is determined 
by the boundary condition of the spin current at the ferromagnetic/nonmagnetic interface, as discussed below. 
The conductivity and spin Hall angle of the nonmagnet are $\sigma_{\rm N}$ and $\vartheta$, respectively. 
The elementary charge is $e=|e|$. 
The Levi-Civita asymmetric tensor is $\epsilon_{ijk}$ ($\epsilon_{123}=+1$). 
The electrochemical potential $\bar{\mu}$ and spin accumulation $\delta\mu$ are 
defined using the spin-dependent electrochemical potential $\bar{\mu}_{\nu}$ ($\nu=\uparrow,\downarrow$) as 
$\bar{\mu}=(\bar{\mu}_{\uparrow}+\bar{\mu}_{\downarrow})/2$ and $\delta\mu=(\bar{\mu}_{\uparrow}-\bar{\mu}_{\downarrow})/2$, respectively. 
The electrochemical potential in the nonmagnet is related to the external electric field as $\bar{\mu}_{\rm N}=eE_{x}x$ [\onlinecite{takahashi08}]. 
On the other hand, the spin accumulation of the nonmagnet varies along the $z$ direction. 
The electric current density along the $z$ direction then becomes zero, guaranteeing the open circuit boundary condition. 
The electric current densities in the nonmagnet 
along the $x$ and $y$ directions are explicitly given by 
\begin{equation}
  J_{{\rm c}x,{\rm N}}
  =
  \sigma_{\rm N}
  E_{x}
  +
  \frac{\vartheta \sigma_{\rm N}}{e}
  \partial_{z}
  \delta
  \mu_{{\rm N},y},
  \label{eq:current_N_x}
\end{equation}
\begin{equation}
  J_{{\rm c}y,{\rm N}}
  =
  -\frac{\vartheta \sigma_{\rm N}}{e}
  \partial_{z}
  \delta
  \mu_{{\rm N},x}.
  \label{eq:current_N_y}
\end{equation}

On the other hand, the electric current density in the ferromagnet is given by [\onlinecite{taniguchi15}] 
\begin{equation}
\begin{split}
  J_{{\rm c}i,{\rm F}}
  =&
  \frac{\sigma_{\rm F}}{e}
  \partial_{i}
  \bar{\mu}_{\rm F}
  +
  \frac{\sigma_{\rm AH}}{e}
  \epsilon_{i \alpha k}
  m_{\alpha}
  \partial_{k}
  \bar{\mu}_{\rm F}
\\
  &
  +
  \beta
  \frac{\sigma_{\rm F}}{e}
  \partial_{i}
  \delta
  \mu_{\rm F}
  +
  \zeta
  \frac{\sigma_{\rm AH}}{e}
  \epsilon_{i \alpha k}
  m_{\alpha}
  \partial_{k}
  \delta
  \mu_{\rm F},
  \label{eq:current_F}
\end{split}
\end{equation}
where $\sigma_{\rm AH}$ is the conductivity of the anomalous Hall effect. 
The spin polarizations of $\sigma_{\rm F}$ and $\sigma_{\rm AH}$ are denoted as $\beta$ and $\zeta$, respectively [\onlinecite{taniguchi15}]. 
The second and fourth terms in Eq. (\ref{eq:current_F}) indicate that 
the anomalous Hall effect generates an electric current perpendicular to both the external current and magnetization. 
We assume that the penetration depth of the transverse spin current in the ferromagnet is sufficiently short 
due to the large exchange interaction between the conduction electrons and the magnetization, for simplicity [\onlinecite{slonczewski96,stiles02,zhang02,zhang04,taniguchi08,ghosh12}]. 
The spin polarizations of the spin current and accumulation in the ferromagnet then become parallel to the magnetization $\mathbf{m}$. 
Using the conservation law of the electric current, $\partial_{i}J_{{\rm c}i,{\rm F}}=0$, 
and applying the open-circuit boundary condition along the $z$ direction, $J_{{\rm c}z,{\rm F}}=0$, 
we find that the electrochemical potential and spin accumulation in the ferromagnet are related as 
\begin{equation}
  \bar{\mu}_{\rm F}
  =
  -\beta
  \delta
  \mu_{\rm F}
  +
  e E_{x} x
  +
  \frac{\sigma_{\rm AH}}{\sigma_{\rm F}}
  m_{y}
  eE_{x}
  z,
  \label{eq:electrochemical_potential_F}
\end{equation}
where the second term is the conventional electric potential, 
whereas the third term arises from an internal electric field under the open-circuit boundary condition. 
The electric current densities in the ferromagnet along the $x$ and $y$ directions are 
\begin{equation}
\begin{split}
  J_{{\rm c}x,{\rm F}}
  =&
  \sigma_{\rm F}
  \left[
    1
    +
    \left(
      \frac{\sigma_{\rm AH}}{\sigma_{\rm F}}
      m_{y}
    \right)^{2}
  \right]
  E_{x}
\\
  &-
  \frac{(\beta-\zeta) \sigma_{\rm AH}}{e}
  m_{y}
  \partial_{z}
  \delta
  \mu_{\rm F}, 
  \label{eq:current_F_x}
\end{split}
\end{equation}
\begin{equation}
\begin{split}
  J_{{\rm c}y,{\rm F}}
  =&
  \sigma_{\rm AH}
  \left[
    m_{z}
    -
    \left(
      \frac{\sigma_{\rm AH}}{\sigma_{\rm F}}
    \right)
    m_{x}
    m_{y}
  \right]
  E_{x}
\\
  &
  +
  \frac{(\beta-\zeta)\sigma_{\rm AH}}{e}
  m_{x}
  \partial_{z}
  \delta
  \mu_{\rm F}.
  \label{eq:current_F_y}
\end{split}
\end{equation}
The last terms in Eqs. (\ref{eq:current_F_x}) and (\ref{eq:current_F_y}) represent 
the contributions to the longitudinal and transverse electric currents 
from the charge-spin conversion caused by the anomalous Hall effect. 
In the following, we will calculate the magnetoresistance effect due to these terms.  


\subsection{Spin current and spin accumulation}

The spin accumulation in the nonmagnet obeys the diffusion equation given by [\onlinecite{valet93,takahashi08}] 
\begin{equation}
  \partial_{z}^{2}
  \delta
  \mu_{{\rm N},\alpha}
  =
  \frac{\delta\mu_{{\rm N},\alpha}}{\ell_{\rm N}^{2}},
  \label{eq:diffusion_equation_N}
\end{equation}
where $\ell_{\rm N}$ is the spin diffusion length of the nonmagnet. 
The boundary conditions of Eq. (\ref{eq:diffusion_equation_N}) is given by the spin current density at the boundary, 
where the spin current density is related to the spin accumulation via [\onlinecite{takahashi08}] 
\begin{equation}
  J_{{\rm s}i\alpha,{\rm N}}
  =
  -\frac{\hbar \sigma_{\rm N}}{2e^{2}}
  \partial_{i}
  \delta
  \mu_{{\rm N},\alpha}
  -
  \frac{\hbar \vartheta \sigma_{\rm N}}{2e^{2}}
  \epsilon_{i \alpha k}
  \partial_{k}
  \bar{\mu}_{\rm N}. 
  \label{eq:spin_current_N}
\end{equation}
To simplify the notation, 
we introduce the unit vectors $\mathbf{e}_{\alpha}$ representing the direction of the spin polarization, 
and define $\delta \bm{\mu}_{\rm N}=\delta \mu_{{\rm N},\alpha}\mathbf{e}_{\alpha}$ and $\mathbf{J}_{{\rm s}z,{\rm N}}=J_{{\rm s}z \alpha,{\rm N}}\mathbf{e}_{\alpha}$. 
The open-circuit boundary condition along the $z$ direction is $\mathbf{J}_{{\rm s}z,{\rm N}}=\bm{0}$ at $z=-d_{\rm N}$. 
On the other hand, we denote the spin current density at the ferromagnetic/nonmagnetic interface ($z=0$) as 
$\mathbf{J}_{{\rm s}z}^{\rm F/N}$. 
The spin accumulation in the nonmagnet is then obtained from Eq. (\ref{eq:diffusion_equation_N}) as 
\begin{equation}
\begin{split}
  \delta
  \bm{\mu}_{{\rm N}}
  =&
  \frac{2e^{2} \ell_{\rm N}}{\hbar \sigma_{\rm N} \sinh(d_{\rm N}/\ell_{\rm N})}
  \left[
    -\frac{\hbar \vartheta \sigma_{\rm N}}{2e}
    E_{x}
    \cosh
    \left(
      \frac{z}{\ell_{\rm N}}
    \right)
    \mathbf{e}_{y}
  \right.
\\
  &
  \left.
    -
    \left(
      \mathbf{J}_{{\rm s}z}^{\rm F/N}
      -
      \frac{\hbar \vartheta \sigma_{\rm N}}{2e}
      E_{x}
      \mathbf{e}_{y}
    \right)
    \cosh
    \left(
      \frac{z+d_{\rm N}}{\ell_{\rm N}}
    \right)
  \right]
  \label{eq:spin_accumulation_N}
\end{split}
\end{equation}
The spin accumulation in the ferromagnet also obeys the diffusion equation 
with the spin diffusion length $\ell_{\rm F}$. 
Note that the spin current density in the ferromagnet is related to the spin accumulation as [\onlinecite{taniguchi15}] 
\begin{equation}
\begin{split}
  J_{{\rm s}i,{\rm F}}
  =&
  -\frac{\hbar \sigma_{\rm F}}{2e^{2}}
  \partial_{i}
  \delta
  \mu_{\rm F}
  -
  \frac{\hbar \sigma_{\rm AH}}{2e^{2}}
  \epsilon_{i \alpha k}
  m_{\alpha}
  \partial_{k}
  \delta
  \mu_{\rm F}
\\
  &-
  \frac{\hbar \beta \sigma_{\rm F}}{2e^{2}}
  \partial_{i}
  \bar{\mu}_{\rm F}
  -
  \frac{\hbar \zeta \sigma_{\rm AH}}{2e^{2}}
  \epsilon_{i \alpha k}
  m_{\alpha}
  \partial_{k}
  \bar{\mu}_{\rm F}.
  \label{eq:spin_current_F}
\end{split}
\end{equation}
Similar to the nonmagnet, we define $\delta\bm{\mu}_{\rm F}=\delta\mu_{\rm F}\mathbf{m}$ 
where the vector notation in boldface represents the direction of the spin polarization, 
which in the ferromagnet is assumed to be parallel to the magnetization. 
Then, the spin accumulation in the ferromagnet is given by 
\begin{equation}
\begin{split}
  \delta
  \mu_{\rm F}
  =&
  \frac{2e^{2}\ell_{\rm F}}{\hbar \sigma_{{\rm F}}^{*} \sinh(d_{\rm F}/\ell_{\rm F})}
  \left\{
    -\frac{\hbar(\beta-\zeta) \sigma_{\rm AH} m_{y}}{2e}
    E_{x}
    \cosh
    \left(
      \frac{z}{\ell_{\rm F}}
    \right)
  \right.
\\
  &
  +
  \left.
    \left[
      \mathbf{m}
      \cdot
      \mathbf{J}_{{\rm s}z}^{\rm F/N}
      +
      \frac{\hbar(\beta-\zeta)\sigma_{\rm AH} m_{y}}{2e}
      E_{x}
    \right]
    \cosh
    \left(
      \frac{z-d_{\rm F}}{\ell_{\rm F}}
    \right)
  \right\}, 
  \label{eq:spin_accumulation_F}
\end{split}
\end{equation}
where $\sigma_{\rm F}^{*}=(1-\beta^{2})\sigma_{\rm F}$. 


\subsection{Spin current at ferromagnetic/nonmagnetic interface}

The spin current at the ferromagnetic/nonmagnetic interface is determined by the spin-dependent Landauer formula 
given by [\onlinecite{brataas01}] 
\begin{equation}
\begin{split}
  \mathbf{J}_{{\rm s}z}^{\rm F/N}
  =&
  -\frac{1}{2\pi S}
  \left[
    \frac{(1-\gamma^{2})g}{2}
    \mathbf{m}
    \cdot
    \left(
      \delta
      \bm{\mu}_{\rm F}
      -
      \delta
      \bm{\mu}_{\rm N}
    \right)
    \mathbf{m}
  \right.
\\
  &
  \left.
    -g_{\rm r}
    \mathbf{m}
    \times
    \left(
      \delta
      \bm{\mu}_{\rm N}
      \times
      \mathbf{m}
    \right)
    -
    g_{\rm i}
    \delta
    \bm{\mu}_{\rm N}
    \times
    \mathbf{m}
  \right],
  \label{eq:spin_current_FN}
\end{split}
\end{equation}
where $g=g^{\uparrow\uparrow}+g^{\downarrow\downarrow}$ is the sum of 
the dimensionless interface conductances of spin-up and spin-down electrons, 
and $\gamma=(g^{\uparrow\uparrow}-g^{\downarrow\downarrow})/g$ is its spin polarization. 
The conductance $g$ is related to the interface resistance $r$ via $r=(h/e^{2})S/g$, 
where $S$ is the cross-section area of the interface. 
The real and imaginary parts of the mixing conductance are denoted as $g_{\rm r}$ and $g_{\rm i}$, respectively. 
Substituting Eqs. (\ref{eq:spin_accumulation_N}) and (\ref{eq:spin_accumulation_F}) at $z=0$ into Eq. (\ref{eq:spin_current_FN}), 
we find that Eq. (\ref{eq:spin_current_FN}) is rewritten as 
\begin{equation}
\begin{split}
  \mathbf{J}_{{\rm s}z}^{\rm F/N}
  =&
  -\frac{\hbar g^{*} m_{y}}{2e}
  \left[
    \frac{(\beta-\zeta) \sigma_{\rm AH}}{g_{\rm F}}
    \tanh
    \left(
      \frac{d_{\rm F}}{2 \ell_{\rm F}}
    \right)
  \right.
\\
   &
   \left.
   -
    \frac{\vartheta \sigma_{\rm N}}{g_{\rm N}}
    \tanh
    \left(
      \frac{d_{\rm N}}{2 \ell_{\rm N}}
    \right)
  \right]
  E_{x}
  \mathbf{m}
\\
  &+
  \frac{\hbar \vartheta \sigma_{\rm N}}{2e}
  \tanh
  \left(
    \frac{d_{\rm N}}{2 \ell_{\rm N}}
  \right)
  E_{x}
  \left[
    \mathbf{m}
    \times
    \left(
      \mathbf{e}_{y}
      \times
      \mathbf{m}
    \right)
    {\rm Re}
  \right.
\\
  &
  \left.
    +
    \mathbf{e}_{y}
    \times
    \mathbf{m}
    {\rm Im}
  \right]
  \frac{g_{\rm r}+ig_{\rm i}}{g_{\rm N} + (g_{\rm r}+ig_{\rm i}) \coth(d_{\rm N}/\ell_{\rm N})}.
  \label{eq:spin_current_FN_sol}
\end{split}
\end{equation}
Here, we define $g_{\rm F}/S=h \sigma_{\rm F}^{*}/(2e^{2}\ell_{\rm F})$, 
$g_{\rm N}/S=h \sigma_{\rm N}/(2e^{2}\ell_{\rm N})$, and 
\begin{equation}
  \frac{1}{g^{*}}
  =
  \frac{2}{(1-\gamma^{2})g}
  +
  \frac{1}{g_{\rm F} \tanh(d_{\rm F}/\ell_{\rm F})}
  +
  \frac{1}{g_{\rm N} \tanh(d_{\rm N}/\ell_{\rm N})}.
\end{equation}
Substituting Eq. (\ref{eq:spin_current_FN_sol}) into Eqs. (\ref{eq:spin_accumulation_N}) and (\ref{eq:spin_accumulation_F}),
the longitudinal and transverse electric current densities 
are calculated from Eqs. (\ref{eq:current_N_x}), (\ref{eq:current_N_y}), (\ref{eq:current_F_x}), and (\ref{eq:current_F_y}). 


\section{Magnetoresistance effect due to charge-spin conversion}
\label{sec:Magnetoresistance effect due to charge-spin conversion}

In this section, we study the magnetoresistance effect 
originated from the charge-spin conversion by the spin Hall and anomalous Hall effects. 
We first focus on the transverse resistivity 
because the transverse voltage has been usually measured in the experiments of the anomalous Hall effect [\onlinecite{pugh53}]. 
We also study the longitudinal resistivity for generality 
because the longitudinal voltage has been measured in the experiments of the spin Hall magnetoresistance [\onlinecite{nakayama13}]. 


\subsection{Transverse resistivity}

We define the averaged transverse electric current density in the nonmagnet from Eq. (\ref{eq:current_N_y}) as 
$\overline{J_{{\rm c}y,{\rm N}}} =\frac{1}{d_{\rm N}} \int_{-d_{\rm N}}^{0} J_{{\rm c}y,{\rm N}} dz$.
Then, the transverse resistivity in the nonmagnet is defined as [\onlinecite{chen13}] 
\begin{equation}
  \rho_{\rm N}^{\rm T}
  \equiv
  -\frac{\overline{J_{{\rm c}y,{\rm N}}}/E_{x}}{\sigma_{\rm N}^{2}}
  =
  \Delta \rho_{{\rm N},y}
  m_{x}
  m_{y}
  +
  \Delta \rho_{{\rm N},y}^{\prime}
  m_{z},
  \label{eq:transverse_resistivity_N}
\end{equation}
where $\Delta \rho_{{\rm N},y}$ and $\Delta\rho_{{\rm N},y}^{\prime}$ with $\rho_{\rm N}=1/\sigma_{\rm N}$ are 
\begin{equation}
\begin{split}
  \frac{\Delta \rho_{{\rm N},y}}{\rho_{\rm N}}
  =&
  \frac{\ell_{\rm N}}{d_{\rm N}}
  \!\!
  \left\{
    \vartheta^{2}
    \!\!
    \left[
      {\rm Re}
      \frac{g_{\rm r} + i g_{\rm i}}{g_{\rm N} + (g_{\rm r} + i g_{\rm i}) \coth(d_{\rm N}/\ell_{\rm N})}
      -
      \frac{g^{*}}{g_{\rm N}}
    \right]
    \!\!
    \tanh
    \!\!
    \left(
      \!
      \frac{d_{\rm N}}{2 \ell_{\rm N}}
      \!
    \right)
  \right.
\\
  &
  \left.
    +
    \frac{\vartheta(\beta-\zeta) \sigma_{\rm AH} g^{*}}{\sigma_{\rm N} g_{\rm F}}
    \tanh
    \left(
      \frac{d_{\rm F}}{2 \ell_{\rm F}}
    \right)
  \right\}
  \tanh
  \left(
    \frac{d_{\rm N}}{2\ell_{\rm N}}
  \right).
  \label{eq:SMR_N}
\end{split}
\end{equation}
\begin{equation}
  \frac{\Delta \rho_{{\rm N},y}^{\prime}}{\rho_{\rm N}}
  =
  -\frac{\vartheta^{2} \ell_{\rm N}}{d_{\rm N}}
  {\rm Im}
  \frac{g_{\rm r} + i g_{\rm i}}{g_{\rm N} + (g_{\rm r} + i g_{\rm i}) \coth(d_{\rm N}/\ell_{\rm N})}
  \tanh^{2}
  \left(
    \frac{d_{\rm N}}{2 \ell_{\rm N}}
  \right). 
\end{equation}
In the following, we neglect terms related to $g_{\rm i}$ by assuming $g_{\rm r} \gg g_{\rm i}$, for simplicity [\onlinecite{zwierzycki05}]. 
Similarly, we define the averaged transverse electric current density in the ferromagnet from Eq. (\ref{eq:current_F_y}) as 
$\overline{J_{{\rm c}y,{\rm F}}} = \frac{1}{d_{\rm F}} \int_{0}^{d_{\rm F}} J_{{\rm c}y,{\rm F}} dz$.
Then, the transverse resistivity in the ferromagnet is 
\begin{equation}
\begin{split}
  \rho_{\rm F}^{\rm T}
  \equiv
  -\frac{\overline{J_{{\rm c}y,{\rm F}}}/E_{x}}{\sigma_{\rm F}^{2}}
  =&
  -\rho_{\rm F}
  \left(
    \frac{\sigma_{\rm AH}}{\sigma_{\rm F}}
  \right)
  m_{z}
  +
  \rho_{\rm F}
  \left(
    \frac{\sigma_{\rm AH}}{\sigma_{\rm F}}
  \right)^{2}
  m_{x}
  m_{y}
\\
  &+
  \Delta
  \rho_{{\rm F},y}
  m_{x}
  m_{y}, 
  \label{eq:transverse_resistivity_F}
\end{split}
\end{equation}
where $\Delta \rho_{{\rm F},y}$ with $\rho_{\rm F}=1/\sigma_{\rm F}$ is 
\begin{equation}
\begin{split}
  \frac{\Delta \rho_{{\rm F},y}}{\rho_{\rm F}}
  =&
  \frac{\ell_{\rm F}}{(1-\beta^{2}) d_{\rm F}}
  \left\{
    \left[
      \frac{(\beta-\zeta) \sigma_{\rm AH}}{\sigma_{\rm F}}
    \right]^{2}
    \left[
      2
      -
      \frac{g^{*}}{g_{\rm F}}
      \tanh
      \left(
        \frac{d_{\rm F}}{2 \ell_{\rm F}}
      \right)
    \right]
  \right.
\\
  &+
  \left.
    \frac{\vartheta(\beta-\zeta) \sigma_{\rm N} \sigma_{\rm AH} g^{*}}{\sigma_{\rm F}^{2} g_{\rm N}}
    \tanh
    \left(
      \frac{d_{\rm N}}{2 \ell_{\rm N}}
    \right)
  \right\}
  \tanh
  \left(
    \frac{d_{\rm F}}{2 \ell_{\rm F}}
  \right).
  \label{eq:SMR_F}
\end{split}
\end{equation}
The first term in Eq. (\ref{eq:transverse_resistivity_F}) is the transverse resistivity due to the conventional anomalous Hall effect, 
whereas the second term is originated from the internal electric field due to the open-circuit condition of the electric current along the $z$ direction. 
On the other hand, the third term or, equivalently, Eq. (\ref{eq:SMR_F}), represents 
the contribution from the charge-spin conversion caused by the anomalous Hall effect. 


Equations (\ref{eq:SMR_N}) and (\ref{eq:SMR_F}) indicate that 
the transverse resistivity shows the angular dependence that is the same as with the planar Hall effect in ferromagnets [\onlinecite{mcguire75}]. 
However, the dependences of Eqs. (\ref{eq:SMR_N}) and (\ref{eq:SMR_F}) on the thicknesses $d_{\rm N}$ and $d_{\rm F}$ are scaled by the spin diffusion lengths, 
and thus these effects can be separated from the conventional planar Hall effect by measuring the thickness dependence of total transverse resistivity. 
The terms in Eqs. (\ref{eq:SMR_N}) and (\ref{eq:SMR_F}) are classified into three contributions. 
The first one is the spin Hall magnetoresistance [\onlinecite{chen13}] described by the first term of Eq. (\ref{eq:SMR_N}). 
This term arises purely from the spin Hall effect, and is finite even when the ferromagnet does not show the anomalous Hall effect. 
The second contribution is the first term of Eq. (\ref{eq:SMR_F}) originating from 
the charge-spin conversion purely caused by the anomalous Hall effect. 
This term is finite even when the spin Hall effect is absent. 
These first and second contributions have fixed signs 
because these are proportional to the square of the spin Hall angle $\vartheta$ and 
the spin polarization $\beta-\zeta$, respectively. 
On the other hand, the third contribution corresponds to the second terms of Eqs. (\ref{eq:SMR_N}) and (\ref{eq:SMR_F}), 
which originate from the charge-spin conversion generated by the mixing of the spin Hall and anomalous Hall effects. 
In other words, the spin current generated by the anomalous Hall effect is converted to the electric current 
by the inverse spin Hall effect or vice versa. 
The sign of the third contribution depends on those of $\vartheta$ and $\beta-\zeta$. 



\begin{figure*}
\centerline{\includegraphics[width=2.0\columnwidth]{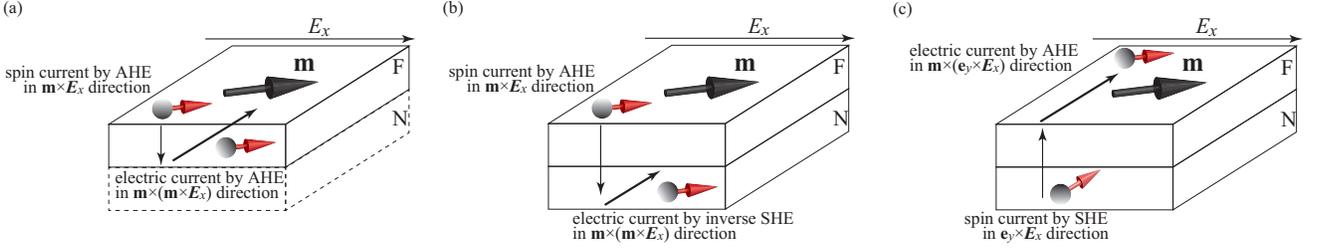}}
\caption{
         Schematic pictures of the generation of the transverse ($y$) electric currents by the charge-spin conversions. 
         (a) The anomalous Hall effect (AHE) generates the spin current in the $\mathbf{m} \times E_{x}\mathbf{e}_{x}$ direction, 
             and converts the spin current into the electric current in the $\mathbf{m} \times (\mathbf{m} \times E_{x} \mathbf{e}_{x})$ direction. 
             The nonmagnet does not play any role in this situation, and thus is represented by the dashed line. 
         (b) The spin current generated by the anomalous Hall effect is converted to the electric current by the inverse spin Hall effect (SHE) in the nonmagnet. 
         (c) The spin current generated by the spin Hall effect is injected into the ferromagnet, and is converted to the electric current by the anomalous Hall effect. 
         \vspace{-3ex}}
\label{fig:fig2}
\end{figure*}




Although all terms in Eqs. (\ref{eq:SMR_N}) and (\ref{eq:SMR_F}) have the angular dependence that is the same as with 
the planar Hall effect, their origins are different. 
The angular dependence of the spin Hall magnetoresistance arises from the absorption of the spin current 
having the spin polarization perpendicular to the magnetization at the ferromagnetic/nonmagnetic interface [\onlinecite{chen13}]. 
Thus, the amount of the spin Hall magnetoresistance depends on the mixing conductance. 
On the other hand, the magnetoresistance due to the anomalous Hall effect arises from the spin current polarized along the direction parallel to the magnetization, 
and thus, is independent of the mixing conductance. 
Let us explain the origins of the magnetoresistances in Eqs. (\ref{eq:SMR_N}) and (\ref{eq:SMR_F}), except the spin Hall magnetoresistance, 
with the help of schematic views in Fig. \ref{fig:fig2}. 
Figure \ref{fig:fig2}(a) schematically shows the generation of the transverse electric current by the charge-spin conversion purely caused by the anomalous Hall effect. 
The anomalous Hall effect generates the spin current in the direction $\mathbf{m} \times E_{x}\mathbf{e}_{x}$. 
Therefore, the spin current flows in the $z$ direction when the magnetization has a finite $y$ component $m_{y}$. 
The anomalous Hall effect then converts this spin current to the electric current 
in the direction perpendicular to both the spin current and the magnetization, 
i.e., $\mathbf{m} \times (\mathbf{m} \times E_{x} \mathbf{e}_{x})$ direction. 
Thus, when the $x$ component of the magnetization $m_{x}$ is also finite, 
the electric current along the $y$ direction is generated in the ferromagnet, as shown in Fig. \ref{fig:fig2}(a). 
This term corresponds to the first term of Eq. (\ref{eq:SMR_F}). 
The spin current generated by the anomalous Hall effect is also converted to the electric current in the nonmagnet by the inverse spin Hall effect, 
as schematically shown in Fig. \ref{fig:fig2}(b), 
which corresponds to the second term in Eq. (\ref{eq:SMR_N}). 
On the other hand, Fig. \ref{fig:fig2}(c) schematically shows the conversion of the spin current generated in the nonmagnet 
to the electric current in the ferromagnet. 
The spin Hall effect generates the spin current in the $z$ direction, whose polarization points to the $y$ direction. 
This spin current can be injected into the ferromagnet when $m_{y}$ is finite 
because only the spin current polarized in the magnetization direction can survive inside the ferromagnet. 
Then, the anomalous Hall effect converts this spin current into the electric current flowing in the direction, 
$\mathbf{m} \times (\mathbf{e}_{y} \times E_{x} \mathbf{e}_{x}) \propto \mathbf{m} \times \mathbf{e}_{z}$. 
Therefore, the electric current in the $y$ direction is generated when $m_{x}$ is finite, 
which corresponds to the second term of Eq. (\ref{eq:SMR_F}). 
These physical pictures indicate that both the $x$ and $y$ components of the magnetization should be finite 
to generate the transverse electric current. 
Therefore, the magnetoresistances in Eqs. (\ref{eq:SMR_N}) and (\ref{eq:SMR_F}) show the angular dependence of $m_{x}m_{y}$. 




\begin{figure*}
\centerline{\includegraphics[width=2.0\columnwidth]{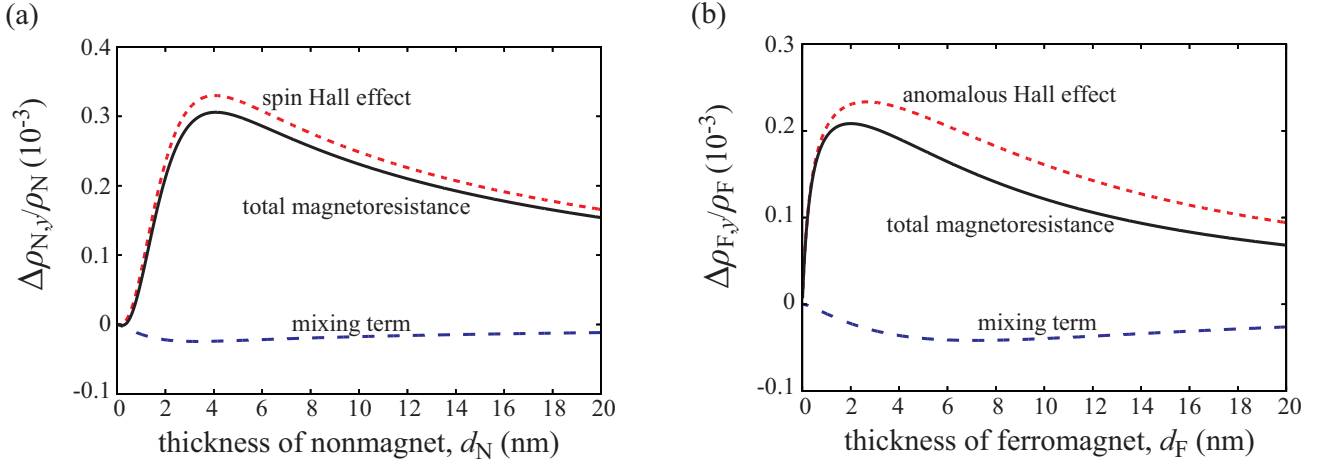}}
\caption{
         (a) Dependence of the transverse resistivity in the nonmagnet normalized by $\rho_{\rm N}$, given by Eq. (\ref{eq:SMR_N}), on the thickness $d_{\rm N}$ (solid line). 
             The dotted (red) and dashed (blue) lines represent the contribution from the spin Hall magnetoresistance and 
             the mixing term of the spin Hall and anomalous Hall effect, 
             which correspond to the first and second term of Eq. (\ref{eq:SMR_N}), respectively. 
         (b) Dependence of the transverse resistivity in the ferromagnet normalized by $\rho_{\rm F}$, given by Eq. (\ref{eq:SMR_F}), on the thickness $d_{\rm F}$ (solid line). 
             Similar to (a), the first and second terms in Eq. (\ref{eq:SMR_F}) are shown by 
             the dotted (red) and dashed (blue) lines, respectively.
             The factors considering the shunt effect are multiplied to both Eqs. (\ref{eq:SMR_N}) and (\ref{eq:SMR_F}). 
         \vspace{-3ex}}
\label{fig:fig3}
\end{figure*}




Figure \ref{fig:fig3}(a) shows the dependence of the transverse resistivity in the nonmagnet, given by Eq. (\ref{eq:SMR_N}), 
on its thickness $d_{\rm N}$ by the solid line. 
A factor $1/[1 + (\rho_{\rm N}d_{\rm F})/(\rho_{\rm F}d_{\rm N})]$ is multiplied to Eq. (\ref{eq:SMR_N}) in this figure 
to account for the shunt effect [\onlinecite{kim16}]. 
The values of the parameters are taken from typical experiments and first-principles calculations 
for CoFeB, NiFe, W, and so on 
[\onlinecite{kim16,taniguchi15,zwierzycki05,bass07,oshima02,moritz08,niimi12}];
$\rho_{\rm F}=300$ $\Omega$nm, $\ell_{\rm F}=5$ nm, $\beta=0.75$, $r=0.25$ k$\Omega$nm${}^{2}$, $\gamma=0.50$, 
$\sigma_{\rm AH}/\sigma_{\rm F}=0.015$, $\zeta=1.5$, $\rho_{\rm N}=1500$ $\Omega$nm, $\ell_{\rm N}=1$ nm, $\vartheta=0.1$, and $g_{\rm r}/S=15$ nm${}^{-2}$. 
The thickness of the ferromagnet is fixed to $d_{\rm F}=2$ nm. 
The first and second terms of Eq. (\ref{eq:SMR_N}) are also shown by the dotted and dashed lines, respectively. 
The transverse resistivity in the ferromagnet, given by Eq. (\ref{eq:SMR_F}), multiplied by the factor $1/[1 + (\rho_{\rm F}d_{\rm N})/(\rho_{\rm N}d_{\rm F})]$, 
as a function of the thickness $d_{\rm F}$ is also shown in Fig. \ref{fig:fig3}(b) 
by a solid line, where $d_{\rm N}$ is fixed to $2$ nm. 
These figures indicate that the magnetoresistance effect on the same order of the spin Hall magnetoresistance can be expected 
due to the charge-spin conversion caused by the anomalous Hall effect. 
It was recently shown that the anisotropic magnetoresistance and the planar Hall effect in the W/CoFeB/MgO heterostructure is 
one order of magnitude smaller than the spin Hall magnetoresistance [\onlinecite{cho15}]. 
Therefore, the magnetoresistance effect studied in the present work is considered to be measurable, 
although it has the same angular dependence with the planar Hall effect. 


\subsection{Longitudinal resistivity}

The longitudinal resistivity in the nonmagnet is defined from Eq. (\ref{eq:current_N_x}) as 
\begin{equation}
  \rho_{\rm N}^{\rm L}
  \equiv
  \left(
    \frac{\overline{J_{{\rm c}x,{\rm N}}}}{E_{x}}
  \right)^{-1}
  \simeq
  \rho_{\rm N}
  +
  \Delta 
  \rho_{{\rm N},x}^{(0)}
  +
  \Delta 
  \rho_{{\rm N},x}^{(1)}
  \left(
    1
    -
    m_{y}^{2}
  \right)
  +
  \Delta
  \rho_{{\rm N},x}^{(2)}
  m_{y}^{2}, 
  \label{eq:longitudinal_resistivity_N}
\end{equation}
where $\Delta\rho_{{\rm N},x}^{(k)}$ ($k=1,2,3$) are, respectively, given by 
\begin{equation}
  \frac{\Delta \rho_{{\rm N},x}^{(0)}}{\rho_{\rm N}}
  =
  -\frac{2 \vartheta^{2} \ell_{\rm N}}{d_{\rm N}}
  \tanh
  \left(
    \frac{d_{\rm N}}{2\ell_{\rm N}}
  \right),
\end{equation}
\begin{equation}
  \frac{\Delta\rho_{{\rm N},x}^{(1)}}{\rho_{\rm N}}
  =
  \frac{\vartheta^{2} \ell_{\rm N}}{d_{\rm N}}
  {\rm Re}
  \frac{g_{\rm r} + i g_{\rm i}}{g_{\rm N}+(g_{\rm r} + i g_{\rm i})\coth(d_{\rm N}/\ell_{\rm N})}
  \tanh^{2}
  \left(
    \frac{d_{\rm N}}{2\ell_{\rm N}}
  \right),
\end{equation}
\begin{equation}
\begin{split}
  \frac{\Delta\rho_{{\rm N},x}^{(2)}}{\rho_{\rm N}}
  =&
  \frac{\ell_{\rm N}}{d_{\rm N}}
  \left[
    \frac{\vartheta^{2} g^{*}}{g_{\rm N}}
    \tanh
    \left(
      \frac{d_{\rm N}}{2 \ell_{\rm N}}
    \right)
  \right.
\\
  &-
  \left.
    \frac{\vartheta(\beta-\zeta) g^{*} \sigma_{\rm AH}}{g_{\rm F} \sigma_{\rm N}}
    \tanh
    \left(
      \frac{d_{\rm F}}{2 \ell_{\rm F}}
    \right)
  \right]
  \tanh
  \left(
    \frac{d_{\rm N}}{2\ell_{\rm N}}
  \right). 
\end{split}
\end{equation}
We note that $\Delta\rho_{{\rm N},x}^{(0)}$ and $\Delta\rho_{{\rm N},x}^{(1)}$ correspond to 
the spin Hall magnetoresistance derived 
for a ferromagnetic insulator/nonmagnetic metal bilayer [\onlinecite{chen13}]. 
The first term in $\Delta\rho_{{\rm N},x}^{(2)}$ is originated from 
the charge-spin conversion purely by the spin Hall effect, 
and is finite when the ferromagnet is metallic. 
On the other hand, the second term in $\Delta\rho_{{\rm N},x}^{(2)}$ is originated from 
the spin current generated by the anomalous Hall effect, 
which is converted to the electric current in the nonmagnet by the inverse spin Hall effect. 


Similarly, the longitudinal resistivity in the ferromagnet is defined from Eq. (\ref{eq:current_F_x}) as  
\begin{equation}
  \rho_{\rm F}^{\rm L}
  \equiv
  \left(
    \frac{\overline{J_{{\rm c}x,{\rm F}}}}{E_{x}}
  \right)^{-1}
  \simeq
  \rho_{\rm F}
  -
  \rho_{\rm F}
  \left(
    \frac{\sigma_{\rm AH}}{\sigma_{\rm F}}
  \right)^{2}
  m_{y}^{2}
  +
  \Delta
  \rho_{{\rm F},x}
  m_{y}^{2}, 
\end{equation}
where $\Delta\rho_{{\rm F},x}$ is given by 
\begin{equation}
\begin{split}
  \frac{\Delta\rho_{{\rm F},x}}{\rho_{\rm F}}
  =&
  -\frac{\ell_{\rm F}}{(1-\beta^{2}) d_{\rm F}}
  \left\{
    \left[
      \frac{(\beta-\zeta) \sigma_{\rm AH}}{\sigma_{\rm F}}
    \right]^{2}
    \left[
      2
      -
      \frac{g^{*}}{g_{\rm F}}
      \tanh
      \left(
        \frac{d_{\rm F}}{2 \ell_{\rm F}}
      \right)
    \right]
  \right.
\\
  &+
  \left.
    \frac{\vartheta(\beta-\zeta) \sigma_{\rm N} \sigma_{\rm AH} g^{*}}{\sigma_{\rm F}^{2} g_{\rm N}}
    \tanh
    \left(
      \frac{d_{\rm N}}{2 \ell_{\rm N}}
    \right)
  \right\}
  \tanh
  \left(
    \frac{d_{\rm F}}{2 \ell_{\rm F}}
  \right).
  \label{eq:longitudinal_resistivity_F}
\end{split}
\end{equation}
As in the case of the transverse resistivity, Eq. (\ref{eq:longitudinal_resistivity_F}) has two terms, 
where the first term arises from the charge-spin conversion purely caused by the anomalous Hall effect [\onlinecite{taniguchi16SPIE}], 
whereas the second term comes from the spin current generated by the spin Hall effect in the nonmagnet. 


We note that the magnitudes of the transverse and longitudinal resistivities in the ferromagnet that originate from the charge-spin conversion, 
given by Eqs. (\ref{eq:SMR_F}) and (\ref{eq:longitudinal_resistivity_F}), are equivalent. 
This is because both the longitudinal and transverse currents due to this conversion, given by the last terms in Eqs. (\ref{eq:current_F_x}) and (\ref{eq:current_F_y}), respectively, 
depend on the same term ($\propto \partial_{z}\delta\mu_{\rm F}$), except the angular dependence. 
Therefore, the dependence of Eq. (\ref{eq:longitudinal_resistivity_F}) on the ferromagnetic thickness is the same with Fig. \ref{fig:fig3}(b), except the sign difference. 
On the other hand, the longitudinal and transverse currents in the nonmagnet given by Eqs. (\ref{eq:current_N_x}) and (\ref{eq:current_N_y}) 
depend on different components of $\delta \mu_{{\rm N},\alpha}$. 
Thus, the magnitudes of the longitudinal and transverse resistivities are, in general, different, 
as can be seen in Eqs. (\ref{eq:SMR_N}) and (\ref{eq:longitudinal_resistivity_N}). 
However, when we compare the longitudinal resistivities for $m_{y}=0$ and $m_{y}=\pm 1$, 
as done in the experiments [\onlinecite{kim16}], 
we found that 
\begin{equation}
  \rho_{\rm N}^{\rm L}(m_{y}=0)
  -
  \rho_{\rm N}^{\rm L}(m_{y}=\pm 1)
  =
  \Delta
  \rho_{{\rm N},x}^{(1)}
  -
  \Delta
  \rho_{{\rm N},x}^{(2)},
\end{equation}
which is identical to the transverse resistivity, given by Eq. (\ref{eq:SMR_N}), and therefore 
its thickness dependence is given by Fig. \ref{fig:fig3}(a). 




\section{Conclusion}
\label{sec:Conclusion}

In conclusion, the magnetoresistance effect in a metallic ferromagnetic/nonmagnetic bilayer due to 
the charge-spin conversion by the anomalous Hall effect is predicted theoretically. 
The magnetoresistance generated from charge-spin conversion purely caused by anomalous Hall effect in the ferromagnet 
is proportional to the square of the spin polarizations in the ferromagnet and has a fixed sign. 
We also find additional magnetoresistances in both the nonmagnet and ferromagnet, 
which arise from the mixing of the spin Hall and anomalous Hall effects. 
The sign of this mixing resistance becomes either positive or negative, 
depending on those of the spin Hall angle in the nonmagnet 
and the spin polarization of the ferromagnet.


\section*{Acknowledgement}

The author expresses gratitude to Mark D. Stiles, Julie Grollier, Takahiko Moriyama, Hitoshi Kubota, and Yoichi Shiota for valuable discussions. 
The author is also thankful to Shinji Yuasa, Akio Fukushima, Kay Yakushiji, Takehiko Yorozu, 
Satoshi Iba, Aurelie Spiesser, Sumito Tsunegi, Atsushi Sugihara, Takahide Kubota, Shinji Miwa, 
Hiroki Maehara, and Ai Emura for their support and encouragement.







\begin{thebibliography}{42}
\expandafter\ifx\csname natexlab\endcsname\relax\def\natexlab#1{#1}\fi
\expandafter\ifx\csname bibnamefont\endcsname\relax
  \def\bibnamefont#1{#1}\fi
\expandafter\ifx\csname bibfnamefont\endcsname\relax
  \def\bibfnamefont#1{#1}\fi
\expandafter\ifx\csname citenamefont\endcsname\relax
  \def\citenamefont#1{#1}\fi
\expandafter\ifx\csname url\endcsname\relax
  \def\url#1{\texttt{#1}}\fi
\expandafter\ifx\csname urlprefix\endcsname\relax\def\urlprefix{URL }\fi
\providecommand{\bibinfo}[2]{#2}
\providecommand{\eprint}[2][]{\url{#2}}

\bibitem[{\citenamefont{Weiler et~al.}(2012)\citenamefont{Weiler, Althammer,
  Czeschka, Huebl, Wagner, Opel, Imort, Reiss, Thomas, Gross
  et~al.}}]{weiler12}
\bibinfo{author}{\bibfnamefont{M.}~\bibnamefont{Weiler}},
  \bibinfo{author}{\bibfnamefont{M.}~\bibnamefont{Althammer}},
  \bibinfo{author}{\bibfnamefont{F.~D.} \bibnamefont{Czeschka}},
  \bibinfo{author}{\bibfnamefont{H.}~\bibnamefont{Huebl}},
  \bibinfo{author}{\bibfnamefont{M.~S.} \bibnamefont{Wagner}},
  \bibinfo{author}{\bibfnamefont{M.}~\bibnamefont{Opel}},
  \bibinfo{author}{\bibfnamefont{I.-M.} \bibnamefont{Imort}},
  \bibinfo{author}{\bibfnamefont{G.}~\bibnamefont{Reiss}},
  \bibinfo{author}{\bibfnamefont{A.}~\bibnamefont{Thomas}},
  \bibinfo{author}{\bibfnamefont{R.}~\bibnamefont{Gross}},
  \bibnamefont{et~al.}, \bibinfo{journal}{Phys. Rev. Lett.}
  \textbf{\bibinfo{volume}{108}}, \bibinfo{pages}{106602}
  (\bibinfo{year}{2012}).

\bibitem[{\citenamefont{Huang et~al.}(2012)\citenamefont{Huang, Fan, Qu, Chen,
  Wang, Wu, Chen, Xiao, and Chien}}]{huang12}
\bibinfo{author}{\bibfnamefont{S.~Y.} \bibnamefont{Huang}},
  \bibinfo{author}{\bibfnamefont{X.}~\bibnamefont{Fan}},
  \bibinfo{author}{\bibfnamefont{D.}~\bibnamefont{Qu}},
  \bibinfo{author}{\bibfnamefont{Y.~P.} \bibnamefont{Chen}},
  \bibinfo{author}{\bibfnamefont{W.~G.} \bibnamefont{Wang}},
  \bibinfo{author}{\bibfnamefont{J.}~\bibnamefont{Wu}},
  \bibinfo{author}{\bibfnamefont{T.~Y.} \bibnamefont{Chen}},
  \bibinfo{author}{\bibfnamefont{J.~Q.} \bibnamefont{Xiao}}, \bibnamefont{and}
  \bibinfo{author}{\bibfnamefont{C.~L.} \bibnamefont{Chien}},
  \bibinfo{journal}{Phys. Rev. Lett.} \textbf{\bibinfo{volume}{109}},
  \bibinfo{pages}{107204} (\bibinfo{year}{2012}).

\bibitem[{\citenamefont{Nakayama et~al.}(2013)\citenamefont{Nakayama,
  Althammer, Chen, Uchida, Kajiwara, Kikuchi, Ohtani, Gepr\"ags, Opel,
  Takahashi et~al.}}]{nakayama13}
\bibinfo{author}{\bibfnamefont{H.}~\bibnamefont{Nakayama}},
  \bibinfo{author}{\bibfnamefont{M.}~\bibnamefont{Althammer}},
  \bibinfo{author}{\bibfnamefont{Y.-T.} \bibnamefont{Chen}},
  \bibinfo{author}{\bibfnamefont{K.}~\bibnamefont{Uchida}},
  \bibinfo{author}{\bibfnamefont{Y.}~\bibnamefont{Kajiwara}},
  \bibinfo{author}{\bibfnamefont{D.}~\bibnamefont{Kikuchi}},
  \bibinfo{author}{\bibfnamefont{T.}~\bibnamefont{Ohtani}},
  \bibinfo{author}{\bibfnamefont{S.}~\bibnamefont{Gepr\"ags}},
  \bibinfo{author}{\bibfnamefont{M.}~\bibnamefont{Opel}},
  \bibinfo{author}{\bibfnamefont{S.}~\bibnamefont{Takahashi}},
  \bibnamefont{et~al.}, \bibinfo{journal}{Phys. Rev. Lett.}
  \textbf{\bibinfo{volume}{110}}, \bibinfo{pages}{206601}
  (\bibinfo{year}{2013}).

\bibitem[{\citenamefont{Althammer et~al.}(2013)\citenamefont{Althammer, Meyer,
  Nakayama, Schreier, Altmannshofer, Weiler, Huebl, Gepr\"ags, Opel, Gross
  et~al.}}]{althammer13}
\bibinfo{author}{\bibfnamefont{M.}~\bibnamefont{Althammer}},
  \bibinfo{author}{\bibfnamefont{S.}~\bibnamefont{Meyer}},
  \bibinfo{author}{\bibfnamefont{H.}~\bibnamefont{Nakayama}},
  \bibinfo{author}{\bibfnamefont{M.}~\bibnamefont{Schreier}},
  \bibinfo{author}{\bibfnamefont{S.}~\bibnamefont{Altmannshofer}},
  \bibinfo{author}{\bibfnamefont{M.}~\bibnamefont{Weiler}},
  \bibinfo{author}{\bibfnamefont{H.}~\bibnamefont{Huebl}},
  \bibinfo{author}{\bibfnamefont{S.}~\bibnamefont{Gepr\"ags}},
  \bibinfo{author}{\bibfnamefont{M.}~\bibnamefont{Opel}},
  \bibinfo{author}{\bibfnamefont{R.}~\bibnamefont{Gross}},
  \bibnamefont{et~al.}, \bibinfo{journal}{Phys. Rev. B}
  \textbf{\bibinfo{volume}{87}}, \bibinfo{pages}{224401}
  (\bibinfo{year}{2013}).

\bibitem[{\citenamefont{Hahn et~al.}(2013)\citenamefont{Hahn, de~Loubens,
  Klein, Viret, Naletov, and BenYoussef}}]{hahn13}
\bibinfo{author}{\bibfnamefont{C.}~\bibnamefont{Hahn}},
  \bibinfo{author}{\bibfnamefont{G.}~\bibnamefont{de~Loubens}},
  \bibinfo{author}{\bibfnamefont{O.}~\bibnamefont{Klein}},
  \bibinfo{author}{\bibfnamefont{M.}~\bibnamefont{Viret}},
  \bibinfo{author}{\bibfnamefont{V.~V.} \bibnamefont{Naletov}},
  \bibnamefont{and}
  \bibinfo{author}{\bibfnamefont{J.}~\bibnamefont{BenYoussef}},
  \bibinfo{journal}{Phys. Rev. B} \textbf{\bibinfo{volume}{87}},
  \bibinfo{pages}{174417} (\bibinfo{year}{2013}).

\bibitem[{\citenamefont{Y.-T.Chen et~al.}(2013)\citenamefont{Y.-T.Chen,
  Takahashi, Nakayama, Althammer, Goennenwein, Saitoh, and Bauer}}]{chen13}
\bibinfo{author}{\bibnamefont{Y.-T.Chen}},
  \bibinfo{author}{\bibfnamefont{S.}~\bibnamefont{Takahashi}},
  \bibinfo{author}{\bibfnamefont{H.}~\bibnamefont{Nakayama}},
  \bibinfo{author}{\bibfnamefont{M.}~\bibnamefont{Althammer}},
  \bibinfo{author}{\bibfnamefont{S.~T.~B.} \bibnamefont{Goennenwein}},
  \bibinfo{author}{\bibfnamefont{E.}~\bibnamefont{Saitoh}}, \bibnamefont{and}
  \bibinfo{author}{\bibfnamefont{G.~E.~W.} \bibnamefont{Bauer}},
  \bibinfo{journal}{Phys. Rev. B} \textbf{\bibinfo{volume}{87}},
  \bibinfo{pages}{144411} (\bibinfo{year}{2013}).

\bibitem[{\citenamefont{Lu et~al.}(2013)\citenamefont{Lu, Cai, Huang, Qu, Miao,
  and Chien}}]{lu13}
\bibinfo{author}{\bibfnamefont{Y.~M.} \bibnamefont{Lu}},
  \bibinfo{author}{\bibfnamefont{J.~W.} \bibnamefont{Cai}},
  \bibinfo{author}{\bibfnamefont{S.~Y.} \bibnamefont{Huang}},
  \bibinfo{author}{\bibfnamefont{D.}~\bibnamefont{Qu}},
  \bibinfo{author}{\bibfnamefont{B.~F.} \bibnamefont{Miao}}, \bibnamefont{and}
  \bibinfo{author}{\bibfnamefont{C.~L.} \bibnamefont{Chien}},
  \bibinfo{journal}{Phys. Rev. B} \textbf{\bibinfo{volume}{87}},
  \bibinfo{pages}{220409} (\bibinfo{year}{2013}).

\bibitem[{\citenamefont{Miao et~al.}(2014)\citenamefont{Miao, Huang, Qu, and
  Chien}}]{miao14}
\bibinfo{author}{\bibfnamefont{B.~F.} \bibnamefont{Miao}},
  \bibinfo{author}{\bibfnamefont{S.~Y.} \bibnamefont{Huang}},
  \bibinfo{author}{\bibfnamefont{D.}~\bibnamefont{Qu}}, \bibnamefont{and}
  \bibinfo{author}{\bibfnamefont{C.~L.} \bibnamefont{Chien}},
  \bibinfo{journal}{Phys. Rev. Lett.} \textbf{\bibinfo{volume}{112}},
  \bibinfo{pages}{236601} (\bibinfo{year}{2014}).

\bibitem[{\citenamefont{Li et~al.}(2014)\citenamefont{Li, Jia, Ding, Liang,
  Luo, and Wu}}]{li14}
\bibinfo{author}{\bibfnamefont{J.~X.} \bibnamefont{Li}},
  \bibinfo{author}{\bibfnamefont{M.~W.} \bibnamefont{Jia}},
  \bibinfo{author}{\bibfnamefont{Z.}~\bibnamefont{Ding}},
  \bibinfo{author}{\bibfnamefont{J.~H.} \bibnamefont{Liang}},
  \bibinfo{author}{\bibfnamefont{Y.~M.} \bibnamefont{Luo}}, \bibnamefont{and}
  \bibinfo{author}{\bibfnamefont{Y.~Z.} \bibnamefont{Wu}},
  \bibinfo{journal}{Phys. Rev. B} \textbf{\bibinfo{volume}{90}},
  \bibinfo{pages}{214415} (\bibinfo{year}{2014}).

\bibitem[{\citenamefont{Zhang}(2000)}]{zhang00}
\bibinfo{author}{\bibfnamefont{S.}~\bibnamefont{Zhang}},
  \bibinfo{journal}{Phys. Rev. Lett.} \textbf{\bibinfo{volume}{85}},
  \bibinfo{pages}{393} (\bibinfo{year}{2000}).

\bibitem[{\citenamefont{Dyakonov}(2007)}]{dyakonov07}
\bibinfo{author}{\bibfnamefont{M.~I.} \bibnamefont{Dyakonov}},
  \bibinfo{journal}{Phys. Rev. Lett.} \textbf{\bibinfo{volume}{99}},
  \bibinfo{pages}{126601} (\bibinfo{year}{2007}).

\bibitem[{\citenamefont{Takahashi and Maekawa}(2008)}]{takahashi08}
\bibinfo{author}{\bibfnamefont{S.}~\bibnamefont{Takahashi}} \bibnamefont{and}
  \bibinfo{author}{\bibfnamefont{S.}~\bibnamefont{Maekawa}},
  \bibinfo{journal}{J. Phys. Soc. Jpn.} \textbf{\bibinfo{volume}{77}},
  \bibinfo{pages}{031009} (\bibinfo{year}{2008}).

\bibitem[{\citenamefont{Ando et~al.}(2008)\citenamefont{Ando, Takahashi, Harii,
  Sasage, Ieda, Maekawa, and Saitoh}}]{ando08}
\bibinfo{author}{\bibfnamefont{K.}~\bibnamefont{Ando}},
  \bibinfo{author}{\bibfnamefont{S.}~\bibnamefont{Takahashi}},
  \bibinfo{author}{\bibfnamefont{K.}~\bibnamefont{Harii}},
  \bibinfo{author}{\bibfnamefont{K.}~\bibnamefont{Sasage}},
  \bibinfo{author}{\bibfnamefont{J.}~\bibnamefont{Ieda}},
  \bibinfo{author}{\bibfnamefont{S.}~\bibnamefont{Maekawa}}, \bibnamefont{and}
  \bibinfo{author}{\bibfnamefont{E.}~\bibnamefont{Saitoh}},
  \bibinfo{journal}{Phys. Rev. Lett.} \textbf{\bibinfo{volume}{101}},
  \bibinfo{pages}{036601} (\bibinfo{year}{2008}).

\bibitem[{\citenamefont{Liu et~al.}(2011)\citenamefont{Liu, Moriyama, Ralph,
  and Buhrman}}]{liu11}
\bibinfo{author}{\bibfnamefont{L.}~\bibnamefont{Liu}},
  \bibinfo{author}{\bibfnamefont{T.}~\bibnamefont{Moriyama}},
  \bibinfo{author}{\bibfnamefont{D.~C.} \bibnamefont{Ralph}}, \bibnamefont{and}
  \bibinfo{author}{\bibfnamefont{R.~A.} \bibnamefont{Buhrman}},
  \bibinfo{journal}{Phys. Rev. Lett.} \textbf{\bibinfo{volume}{106}},
  \bibinfo{pages}{036601} (\bibinfo{year}{2011}).

\bibitem[{\citenamefont{Liu et~al.}(2012)\citenamefont{Liu, Lee, Gudmundsen,
  Ralph, and Buhrman}}]{liu12}
\bibinfo{author}{\bibfnamefont{L.}~\bibnamefont{Liu}},
  \bibinfo{author}{\bibfnamefont{O.~J.} \bibnamefont{Lee}},
  \bibinfo{author}{\bibfnamefont{T.~J.} \bibnamefont{Gudmundsen}},
  \bibinfo{author}{\bibfnamefont{D.~C.} \bibnamefont{Ralph}}, \bibnamefont{and}
  \bibinfo{author}{\bibfnamefont{R.~A.} \bibnamefont{Buhrman}},
  \bibinfo{journal}{Phys. Rev. Lett.} \textbf{\bibinfo{volume}{109}},
  \bibinfo{pages}{096602} (\bibinfo{year}{2012}).

\bibitem[{\citenamefont{Haney et~al.}(2013)\citenamefont{Haney, Lee, Lee,
  Manchon, and Stiles}}]{haney13}
\bibinfo{author}{\bibfnamefont{P.~M.} \bibnamefont{Haney}},
  \bibinfo{author}{\bibfnamefont{H.-W.} \bibnamefont{Lee}},
  \bibinfo{author}{\bibfnamefont{K.-J.} \bibnamefont{Lee}},
  \bibinfo{author}{\bibfnamefont{A.}~\bibnamefont{Manchon}}, \bibnamefont{and}
  \bibinfo{author}{\bibfnamefont{M.~D.} \bibnamefont{Stiles}},
  \bibinfo{journal}{Phys. Rev. B} \textbf{\bibinfo{volume}{87}},
  \bibinfo{pages}{174411} (\bibinfo{year}{2013}).

\bibitem[{\citenamefont{Saslow}(2015)}]{saslow15}
\bibinfo{author}{\bibfnamefont{W.~M.} \bibnamefont{Saslow}},
  \bibinfo{journal}{Phys. Rev. B} \textbf{\bibinfo{volume}{91}},
  \bibinfo{pages}{014401} (\bibinfo{year}{2015}).

\bibitem[{\citenamefont{Dyakonov and Perel}(1971)}]{dyakonov71}
\bibinfo{author}{\bibfnamefont{M.~I.} \bibnamefont{Dyakonov}} \bibnamefont{and}
  \bibinfo{author}{\bibfnamefont{V.~I.} \bibnamefont{Perel}},
  \bibinfo{journal}{Phys. Lett. A} \textbf{\bibinfo{volume}{35}},
  \bibinfo{pages}{459} (\bibinfo{year}{1971}).

\bibitem[{\citenamefont{Hirsch}(1999)}]{hirsch99}
\bibinfo{author}{\bibfnamefont{J.~E.} \bibnamefont{Hirsch}},
  \bibinfo{journal}{Phys. Rev. Lett.} \textbf{\bibinfo{volume}{83}},
  \bibinfo{pages}{1834} (\bibinfo{year}{1999}).

\bibitem[{\citenamefont{Hoffmann}(2013)}]{hoffmann13}
\bibinfo{author}{\bibfnamefont{A.}~\bibnamefont{Hoffmann}},
  \bibinfo{journal}{IEEE Trans. Magn.} \textbf{\bibinfo{volume}{49}},
  \bibinfo{pages}{5172} (\bibinfo{year}{2013}).

\bibitem[{\citenamefont{Avci et~al.}(2015)\citenamefont{Avci, Garello, Ghosh,
  Gabureac, Alvarado, and Gambardella}}]{avci15}
\bibinfo{author}{\bibfnamefont{C.~O.} \bibnamefont{Avci}},
  \bibinfo{author}{\bibfnamefont{K.}~\bibnamefont{Garello}},
  \bibinfo{author}{\bibfnamefont{A.}~\bibnamefont{Ghosh}},
  \bibinfo{author}{\bibfnamefont{M.}~\bibnamefont{Gabureac}},
  \bibinfo{author}{\bibfnamefont{S.~F.} \bibnamefont{Alvarado}},
  \bibnamefont{and}
  \bibinfo{author}{\bibfnamefont{P.}~\bibnamefont{Gambardella}},
  \bibinfo{journal}{Nat. Phys.} \textbf{\bibinfo{volume}{11}},
  \bibinfo{pages}{570} (\bibinfo{year}{2015}).

\bibitem[{\citenamefont{Liu et~al.}(2015)\citenamefont{Liu, Ohkubo, Mitani,
  Hono, and Hayashi}}]{liu15}
\bibinfo{author}{\bibfnamefont{J.}~\bibnamefont{Liu}},
  \bibinfo{author}{\bibfnamefont{T.}~\bibnamefont{Ohkubo}},
  \bibinfo{author}{\bibfnamefont{S.}~\bibnamefont{Mitani}},
  \bibinfo{author}{\bibfnamefont{K.}~\bibnamefont{Hono}}, \bibnamefont{and}
  \bibinfo{author}{\bibfnamefont{M.}~\bibnamefont{Hayashi}},
  \bibinfo{journal}{Appl. Phys. Lett.} \textbf{\bibinfo{volume}{107}},
  \bibinfo{pages}{232408} (\bibinfo{year}{2015}).

\bibitem[{\citenamefont{Cho et~al.}(2015)\citenamefont{Cho, Baek, Lee, Jo, and
  Park}}]{cho15}
\bibinfo{author}{\bibfnamefont{S.}~\bibnamefont{Cho}},
  \bibinfo{author}{\bibfnamefont{S.-H.~C.} \bibnamefont{Baek}},
  \bibinfo{author}{\bibfnamefont{K.-D.} \bibnamefont{Lee}},
  \bibinfo{author}{\bibfnamefont{Y.}~\bibnamefont{Jo}}, \bibnamefont{and}
  \bibinfo{author}{\bibfnamefont{B.-G.} \bibnamefont{Park}},
  \bibinfo{journal}{Sci. Rep.} \textbf{\bibinfo{volume}{5}},
  \bibinfo{pages}{14668} (\bibinfo{year}{2015}).

\bibitem[{\citenamefont{Kim et~al.}(2016)\citenamefont{Kim, Sheng, Takahashi,
  Mitani, and Hayashi}}]{kim16}
\bibinfo{author}{\bibfnamefont{J.}~\bibnamefont{Kim}},
  \bibinfo{author}{\bibfnamefont{P.}~\bibnamefont{Sheng}},
  \bibinfo{author}{\bibfnamefont{S.}~\bibnamefont{Takahashi}},
  \bibinfo{author}{\bibfnamefont{S.}~\bibnamefont{Mitani}}, \bibnamefont{and}
  \bibinfo{author}{\bibfnamefont{M.}~\bibnamefont{Hayashi}},
  \bibinfo{journal}{Phys. Rev. Lett.} \textbf{\bibinfo{volume}{116}},
  \bibinfo{pages}{097201} (\bibinfo{year}{2016}).

\bibitem[{\citenamefont{Pugh and Postoker}(1953)}]{pugh53}
\bibinfo{author}{\bibfnamefont{E.~M.} \bibnamefont{Pugh}} \bibnamefont{and}
  \bibinfo{author}{\bibfnamefont{N.}~\bibnamefont{Postoker}},
  \bibinfo{journal}{Rev. Mod. Phys.} \textbf{\bibinfo{volume}{25}},
  \bibinfo{pages}{151} (\bibinfo{year}{1953}).

\bibitem[{\citenamefont{Miao et~al.}(2013)\citenamefont{Miao, Huang, Qu, and
  Chien}}]{miao13}
\bibinfo{author}{\bibfnamefont{B.~F.} \bibnamefont{Miao}},
  \bibinfo{author}{\bibfnamefont{S.~Y.} \bibnamefont{Huang}},
  \bibinfo{author}{\bibfnamefont{D.}~\bibnamefont{Qu}}, \bibnamefont{and}
  \bibinfo{author}{\bibfnamefont{C.~L.} \bibnamefont{Chien}},
  \bibinfo{journal}{Phys. Rev. Lett.} \textbf{\bibinfo{volume}{111}},
  \bibinfo{pages}{066602} (\bibinfo{year}{2013}).

\bibitem[{\citenamefont{Taniguchi et~al.}(2015)\citenamefont{Taniguchi,
  Grollier, and Stiles}}]{taniguchi15}
\bibinfo{author}{\bibfnamefont{T.}~\bibnamefont{Taniguchi}},
  \bibinfo{author}{\bibfnamefont{J.}~\bibnamefont{Grollier}}, \bibnamefont{and}
  \bibinfo{author}{\bibfnamefont{M.~D.} \bibnamefont{Stiles}},
  \bibinfo{journal}{Phys. Rev. Applied} \textbf{\bibinfo{volume}{3}},
  \bibinfo{pages}{044001} (\bibinfo{year}{2015}).

\bibitem[{\citenamefont{Slonczewski}(1996)}]{slonczewski96}
\bibinfo{author}{\bibfnamefont{J.~C.} \bibnamefont{Slonczewski}},
  \bibinfo{journal}{J. Magn. Magn. Mater.} \textbf{\bibinfo{volume}{159}},
  \bibinfo{pages}{L1} (\bibinfo{year}{1996}).

\bibitem[{\citenamefont{Stiles and Zangwill}(2002)}]{stiles02}
\bibinfo{author}{\bibfnamefont{M.~D.} \bibnamefont{Stiles}} \bibnamefont{and}
  \bibinfo{author}{\bibfnamefont{A.}~\bibnamefont{Zangwill}},
  \bibinfo{journal}{Phys. Rev. B} \textbf{\bibinfo{volume}{66}},
  \bibinfo{pages}{014407} (\bibinfo{year}{2002}).

\bibitem[{\citenamefont{Zhang et~al.}(2002)\citenamefont{Zhang, Levy, and
  Fert}}]{zhang02}
\bibinfo{author}{\bibfnamefont{S.}~\bibnamefont{Zhang}},
  \bibinfo{author}{\bibfnamefont{P.~M.} \bibnamefont{Levy}}, \bibnamefont{and}
  \bibinfo{author}{\bibfnamefont{A.}~\bibnamefont{Fert}},
  \bibinfo{journal}{Phys. Rev. Lett.} \textbf{\bibinfo{volume}{88}},
  \bibinfo{pages}{236601} (\bibinfo{year}{2002}).

\bibitem[{\citenamefont{Zhang et~al.}(2004)\citenamefont{Zhang, Levy, Zhang,
  and Antropov}}]{zhang04}
\bibinfo{author}{\bibfnamefont{J.}~\bibnamefont{Zhang}},
  \bibinfo{author}{\bibfnamefont{P.~M.} \bibnamefont{Levy}},
  \bibinfo{author}{\bibfnamefont{S.}~\bibnamefont{Zhang}}, \bibnamefont{and}
  \bibinfo{author}{\bibfnamefont{V.}~\bibnamefont{Antropov}},
  \bibinfo{journal}{Phys. Rev. Lett.} \textbf{\bibinfo{volume}{93}},
  \bibinfo{pages}{256602} (\bibinfo{year}{2004}).

\bibitem[{\citenamefont{Taniguchi et~al.}(2008)\citenamefont{Taniguchi, Yakata,
  Imamura, and Ando}}]{taniguchi08}
\bibinfo{author}{\bibfnamefont{T.}~\bibnamefont{Taniguchi}},
  \bibinfo{author}{\bibfnamefont{S.}~\bibnamefont{Yakata}},
  \bibinfo{author}{\bibfnamefont{H.}~\bibnamefont{Imamura}}, \bibnamefont{and}
  \bibinfo{author}{\bibfnamefont{Y.}~\bibnamefont{Ando}},
  \bibinfo{journal}{Appl. Phys. Express} \textbf{\bibinfo{volume}{1}},
  \bibinfo{pages}{031302} (\bibinfo{year}{2008}).

\bibitem[{\citenamefont{Ghosh et~al.}(2012)\citenamefont{Ghosh, Auffret, Ebels,
  and Bailey}}]{ghosh12}
\bibinfo{author}{\bibfnamefont{A.}~\bibnamefont{Ghosh}},
  \bibinfo{author}{\bibfnamefont{S.}~\bibnamefont{Auffret}},
  \bibinfo{author}{\bibfnamefont{U.}~\bibnamefont{Ebels}}, \bibnamefont{and}
  \bibinfo{author}{\bibfnamefont{W.~E.} \bibnamefont{Bailey}},
  \bibinfo{journal}{Phys. Rev. Lett.} \textbf{\bibinfo{volume}{109}},
  \bibinfo{pages}{127202} (\bibinfo{year}{2012}).

\bibitem[{\citenamefont{Valet and Fert}(1993)}]{valet93}
\bibinfo{author}{\bibfnamefont{T.}~\bibnamefont{Valet}} \bibnamefont{and}
  \bibinfo{author}{\bibfnamefont{A.}~\bibnamefont{Fert}},
  \bibinfo{journal}{Phys. Rev. B} \textbf{\bibinfo{volume}{48}},
  \bibinfo{pages}{7099} (\bibinfo{year}{1993}).

\bibitem[{\citenamefont{Brataas et~al.}(2001)\citenamefont{Brataas, Nazarov,
  and Bauer}}]{brataas01}
\bibinfo{author}{\bibfnamefont{A.}~\bibnamefont{Brataas}},
  \bibinfo{author}{\bibfnamefont{Y.~V.} \bibnamefont{Nazarov}},
  \bibnamefont{and} \bibinfo{author}{\bibfnamefont{G.~E.~W.}
  \bibnamefont{Bauer}}, \bibinfo{journal}{Eur. Phys. J. B}
  \textbf{\bibinfo{volume}{22}}, \bibinfo{pages}{99} (\bibinfo{year}{2001}).

\bibitem[{\citenamefont{Zwierzycki et~al.}(2005)\citenamefont{Zwierzycki,
  Tserkovnyak, Kelly, Brataas, and Bauer}}]{zwierzycki05}
\bibinfo{author}{\bibfnamefont{M.}~\bibnamefont{Zwierzycki}},
  \bibinfo{author}{\bibfnamefont{Y.}~\bibnamefont{Tserkovnyak}},
  \bibinfo{author}{\bibfnamefont{P.~J.} \bibnamefont{Kelly}},
  \bibinfo{author}{\bibfnamefont{A.}~\bibnamefont{Brataas}}, \bibnamefont{and}
  \bibinfo{author}{\bibfnamefont{G.~E.~W.} \bibnamefont{Bauer}},
  \bibinfo{journal}{Phys. Rev. B} \textbf{\bibinfo{volume}{71}},
  \bibinfo{pages}{064420} (\bibinfo{year}{2005}).

\bibitem[{\citenamefont{McGuire and Potter}(1975)}]{mcguire75}
\bibinfo{author}{\bibfnamefont{T.~R.} \bibnamefont{McGuire}} \bibnamefont{and}
  \bibinfo{author}{\bibfnamefont{R.}~\bibnamefont{Potter}},
  \bibinfo{journal}{IEEE Trans. Magn.} \textbf{\bibinfo{volume}{11}},
  \bibinfo{pages}{1018} (\bibinfo{year}{1975}).

\bibitem[{\citenamefont{Bass and W.~P.~Pratt}(2007)}]{bass07}
\bibinfo{author}{\bibfnamefont{J.}~\bibnamefont{Bass}} \bibnamefont{and}
  \bibinfo{author}{\bibfnamefont{J.}~\bibnamefont{W.~P.~Pratt}},
  \bibinfo{journal}{J. Phys.: Condens. Matter} \textbf{\bibinfo{volume}{19}},
  \bibinfo{pages}{183201} (\bibinfo{year}{2007}).

\bibitem[{\citenamefont{Oshima et~al.}(2002)\citenamefont{Oshima, Nagasaka,
  Seyama, Shimizu, Eguchi, and Tanaka}}]{oshima02}
\bibinfo{author}{\bibfnamefont{H.}~\bibnamefont{Oshima}},
  \bibinfo{author}{\bibfnamefont{K.}~\bibnamefont{Nagasaka}},
  \bibinfo{author}{\bibfnamefont{Y.}~\bibnamefont{Seyama}},
  \bibinfo{author}{\bibfnamefont{Y.}~\bibnamefont{Shimizu}},
  \bibinfo{author}{\bibfnamefont{S.}~\bibnamefont{Eguchi}}, \bibnamefont{and}
  \bibinfo{author}{\bibfnamefont{A.}~\bibnamefont{Tanaka}},
  \bibinfo{journal}{J. Appl. Phys.} \textbf{\bibinfo{volume}{91}},
  \bibinfo{pages}{8105} (\bibinfo{year}{2002}).

\bibitem[{\citenamefont{Moritz et~al.}(2008)\citenamefont{Moritz, Rodmacq,
  Auffret, and Dieny}}]{moritz08}
\bibinfo{author}{\bibfnamefont{J.}~\bibnamefont{Moritz}},
  \bibinfo{author}{\bibfnamefont{B.}~\bibnamefont{Rodmacq}},
  \bibinfo{author}{\bibfnamefont{S.}~\bibnamefont{Auffret}}, \bibnamefont{and}
  \bibinfo{author}{\bibfnamefont{B.}~\bibnamefont{Dieny}}, \bibinfo{journal}{J.
  Phys. D} \textbf{\bibinfo{volume}{41}}, \bibinfo{pages}{135001}
  (\bibinfo{year}{2008}).

\bibitem[{\citenamefont{Niimi et~al.}(2012)\citenamefont{Niimi, Kawanishi, Wei,
  Deranlot, Yang, Chshiev, Valet, Fert, and Otani}}]{niimi12}
\bibinfo{author}{\bibfnamefont{Y.}~\bibnamefont{Niimi}},
  \bibinfo{author}{\bibfnamefont{Y.}~\bibnamefont{Kawanishi}},
  \bibinfo{author}{\bibfnamefont{D.~H.} \bibnamefont{Wei}},
  \bibinfo{author}{\bibfnamefont{C.}~\bibnamefont{Deranlot}},
  \bibinfo{author}{\bibfnamefont{H.~X.} \bibnamefont{Yang}},
  \bibinfo{author}{\bibfnamefont{M.}~\bibnamefont{Chshiev}},
  \bibinfo{author}{\bibfnamefont{T.}~\bibnamefont{Valet}},
  \bibinfo{author}{\bibfnamefont{A.}~\bibnamefont{Fert}}, \bibnamefont{and}
  \bibinfo{author}{\bibfnamefont{Y.}~\bibnamefont{Otani}},
  \bibinfo{journal}{Phys. Rev. Lett.} \textbf{\bibinfo{volume}{109}},
  \bibinfo{pages}{156602} (\bibinfo{year}{2012}).

\bibitem[{\citenamefont{Taniguchi et~al.}(2016)\citenamefont{Taniguchi,
  Grollier, and Stiles}}]{taniguchi16SPIE}
\bibinfo{author}{\bibfnamefont{T.}~\bibnamefont{Taniguchi}},
  \bibinfo{author}{\bibfnamefont{J.}~\bibnamefont{Grollier}}, \bibnamefont{and}
  \bibinfo{author}{\bibfnamefont{M.~D.} \bibnamefont{Stiles}},
  \bibinfo{journal}{Proc. SPIE} \textbf{\bibinfo{volume}{9931}},
  \bibinfo{pages}{99310W} (\bibinfo{year}{2016}).

\end{thebibliography}


\end{document}